\documentclass{aa}
\usepackage{graphics}
\usepackage{epsfig}

\newcommand{\ltsima} {$\; \buildrel < \over \sim \;$}
\newcommand{\gtsima} {$\; \buildrel > \over \sim \;$}
\newcommand{\lta} {\lower.5ex\hbox{\ltsima}}
\newcommand{\gta} {\lower.5ex\hbox{\gtsima}}
\begin{document}

%\setcounter{table}{0}

%\thesaurus{03}
\title{HST images of B2 radio galaxies: a link between circum-nuclear dust and radio properties?
\thanks{Based on observations with the NASA/ESA Hubble Space Telescope,
obtained at the
Space Telescope Science Institute, which is operated by AURA, Inc.,
under NASA contract NAS 5-26555 and by STScI grant GO-3594.01-91A}}
\author{H.R. de Ruiter\inst{1,2} \and P. Parma\inst{2} \and A. Capetti\inst{3}
\and R. Fanti\inst{4,2} \and R. Morganti\inst{5}
}
\institute{Osservatorio Astronomico di Bologna, Via Ranzani, 1, I-40127 Italy
\and Istituto di Radioastronomia, Via Gobetti 101, I-40129, Bologna, Italy
\and Osservatorio Astronomico di Torino, Strada Osservatorio 25,  I-10025 Pino Torinese, Italy
\and Istituto di Fisica, University of Bologna, Via Irnerio 46, I-40126 Bologna, Italy
\and Netherlands Foundation for Research in Astronomy, Postbus 2, NL-7990 AA, Dwingeloo,
The Netherlands
}
\offprints{H.R.\, de Ruiter}
\date{Received 4 May 2002; accepted 1 October 2002}
\titlerunning{Dust in low luminosity radio galaxies}
\authorrunning{de Ruiter et al.}
\maketitle

\begin{abstract}
\label{abstract}
Almost 60 \% of the B2 low luminosity radio galaxies have been observed with the Hubble Space
Telescope. We present an analysis of the dust features, which are often present in the form of
circum-nuclear disks or lanes, and show that there are correlations between radio source and dust
properties.
It is found that nearby radio sources in which a jet has been detected tend to have dust more
often than sources without jets; the dust is often in the form of disks or lanes. Moreover the
radio jets are close to perpendicular to the disk or lane in the weaker radio sources
(with  $P < 10^{24}$ WHz$^{-1}$). In stronger sources the orientation effect appears to be
weak or even
absent. Also the dust masses found in the weaker radio sources are smaller than in the stronger
ones ($\log M/M_{\sun} \sim 3$ against 5 respectively). More generally it appears that
there is a correlation between dust mass and total radio power (for those sources in which
% CHANGE: left out "linear"
dust has been detected); we show that this correlation is not induced by redshift. 

\end{abstract}
\section{Introduction}
\label{sec:intro}
A long standing question in radio astronomy has been why some elliptical galaxies host strong
radio sources and others not. The reason may have something to do with the physical conditions near
the active nucleus from which the radio source emanates. Although attempts have been made in the
past to link the presence of optical features like dust lanes and a radio source these have
always been inconclusive due to the lack of adequate angular resolution especially in the optical
images.
The quality of optical imaging has now dramatically improved: imaging with the Hubble Space
Telescope (HST) has provided valuable new information on the optical structure near the nucleus of
radio galaxies. In particular, the HST/WFPC2 snapshot survey of the 3CR sample has produced a large
and uniform database of images essential for a statistical analysis of their host galaxies
(Martel et al. \cite{martel99}, De Koff et al. \cite{dekoff96}). These studies revealed the
presence of new and interesting features,  some of them almost exclusively associated to low
luminosity  FR I radio galaxies.
For example, the HST observations have shown the presence of dust in a large fraction of weak (FR I)
radio galaxies which takes the form of extended nuclear disks (Jaffe et al. \cite{jaffe},
De Koff et al. \cite{dekoff96}, De Juan, Colina \& Golombek \cite{dejuan}, Verdoes Kleijn et al.
\cite{verdoes}). Such structures have been naturally
identified with the reservoir of material which will ultimately accrete into the central black hole.
The symmetry axis of the nuclear disk may be a useful indicator of the the rotation axis of the
central black hole (see e.g. Capetti \& Celotti \cite{capetti99}), although it must be admitted that
the precise relationship between these two axes remains uncertain.
%Several new optical jets have also been found in the HST images (e.g. Sparks et al. \cite{sparks},
%Baum et al. \cite{baum},
%Martel et al. \cite{martel98}): according to Martel et al. (\cite{martel99}) they are seen in
%$\sim13 $% of  nearby($z<0.1$)
% 3CR radio galaxies and, with the only exception of 3C 273, are all found in FR~I radio-galaxies.
FR I sources represent an essential ingredient in the Unified Schemes of radio-loud AGNs as the
mis-oriented counter part of BL Lac objects (see Urry and Padovani \cite{urry95} for a review).
Indeed, recent radio studies have provided strong evidence in favour of relativistic beaming in
their jets (Laing et al. \cite{laing99}). Host galaxies of BL Lacs have been studied with the HST
by Urry et al. (\cite{urry99}), and a similar analysis of galaxies hosting FR I sources should
clarify if there is continuity in the properties of the respective nuclear regions, i.e. if any
differences are due solely to orientation  effects. Furthermore, it was shown by Chiaberge,
Capetti and Celotti (\cite{chiaberge99}, \cite{chiaberge00}) that the optical nuclear sources commonly
found in FR~I might represent the counterpart of the synchrotron radio cores. If this
picture is correct, one expects that the dust features found in FR I sources may provide further
indications on the orientation of the radio source, provided that these features have something to
do with the radio properties.
As many of the most extensively studied FR I radio galaxies are part of the B2 sample of low
luminosity radio galaxies (Fanti et al. \cite{fanti87}), this prompted us to perform a complete,
high resolution optical study of these $\sim 100$ sources. The radio characteristics of the B2 sample
are different and in some sense complementary to those of the 3CR already studied with HST
and they can fill the gap between  the "radio quiet  \& normal"  and the "radio loud" ellipticals.
Although only part (almost 60 \%) of the B2 sample has been observed with the HST, the high
resolution optical data can be considered as representative and unbiased with respect to the
entire B2 sample.  The new HST data were presented in Capetti et al. (\cite{capetti00}).
A statistical study of such a large sample of low luminosity radio galaxies enables us to
establish how frequently we can detect optical jets, optical nuclear sources, circum-nuclear
disks  and dust lanes and if there is any relationship with the radio properties.
The organization of this paper is as follows: in  Sect. \ref{sec:thesample} we briefly describe
the HST data of the B2 sample and give some general information; in Sect. \ref{sec:presenceofdust}
we describe how we produced images in which the properties of the circum-nuclear dust can be
analyzed; a discussion of the dust properties in given in Sect. \ref{sec:discussion}.
Throughout this paper we use a Hubble constant $H_0 = 100$ km s$^{-1}$Mpc$^{-1}$ and $q_0=0.5$.

\section{HST observations of the the Sample}
\label{sec:thesample}

A brief description of the B2 sample of low luminosity radio galaxies and its present status was
given in Capetti et al. (\cite{capetti00}). The B2 sample of low luminosity radio galaxies has been
extensively studied at radio wavelengths, especially since the 1980's (see Fanti et al.
\cite{fanti87}, Parma et al. \cite{parma}, De Ruiter et al. \cite{deruiter},
Morganti et al. \cite{morg97}). Comparison with Einstein Satellite X-ray data was done by
Morganti et al. (\cite{morg88}). A number of B2 radio galaxies were subsequently observed with
ROSAT (Feretti et al. \cite{feretti}, Massaglia et al. \cite{massaglia}, Trussoni et al.
\cite{trussoni}, Canosa et al. \cite{canosa99}). 

Optical work on the entire B2 sample has somewhat lagged behind, but a complete
broad band imaging survey of the B2 sample was carried out by Gonzalez-Serrano et al.
(\cite{gonz93}) and Gonzalez-Serrano \& Carballo (\cite{gonz00}), while narrow band Ha images
were obtained by Morganti et al. (\cite{morg92}). The IRAS properties of the sample were studied
by Impey \& Gregorini (\cite{impey}).

In the course of our HST program observations were obtained for 41 B2 galaxies, while public
archive data exist for 16 additional objects, usually because these sources are also part of
the 3C catalog (see De Koff et al. \cite{dekoff96}, Martel et al. \cite{martel99}, Verdoes Kleijn et al.
\cite{verdoes}). Therefore up till the present HST imaging has been done for 57 of the ~100 radio
galaxies. In the following we will speak loosely of the B2 sample, when we discuss the sample
of these 57 radio galaxies, while in the following the term 3C sample refers to 3C radio galaxies
used by de Koff et al. (\cite{dekoff00}) that are {\it not} part of the B2 sample.

Our program observations were obtained between April 9th and September 1st 1999, using the Wide
Field and Planetary Camera 2 (WFPC2). The pixel size of the Planetary Camera, in which all targets
are located, is 0.0455 arcsec and the $800\times 800$ pixels cover a field of view of
$36\times 36$ arcsec. Two broad
band filters were used, namely F555W and F814W, which cover the spectral regions $4500-6500$ \AA\ and
$7000-9500$ \AA\ respectively. Although their transmission curves do not match exactly those of
standard filters we will usually refer to them as $V$ and $I$ filters. The data have been processed
through the PODPS (Post Observation Data Processing System) pipeline for bias removal and flat
fielding (Biretta et al. \cite{biretta}).
One image was taken through each filter and the exposure time was always set to 300 s
(with the exception of B2 2116+26, which was observed for 350 s in the $V$ band, as $I$ band images
were already available in the archive).

\section{Circum-nuclear dust in low luminosity radio galaxies}
\label{sec:presenceofdust}
In order to study the dust in detail, we made use of the fact that we have both $V$ and $I$ images
(at least for the sources that were observed in our program). Since the galaxy color $V-I$ has only
very small gradients the images in $V$ and $I$ are very similar and differ only by a constant. However,
the dust features are more prominent in the $V$ image, and this suggests we can fit the galaxy in the
$I$ image by a regular elliptical distribution and use this, to a first approximation, as a dust-free
model.  Taking into account the constant offset between  $V$ and $I$, the $V$ image was divided 
by the galaxy model as obtained from the $I$ image. The resulting image of ratios will clearly show any dust features
and thus can be easily analyzed; in regions without absorption the pixel values 
will be close to unity, while absorption results in values $<1$.

\begin{figure}
%\picplace{6.0cm}
%\vspace{8cm}
\epsfig{file=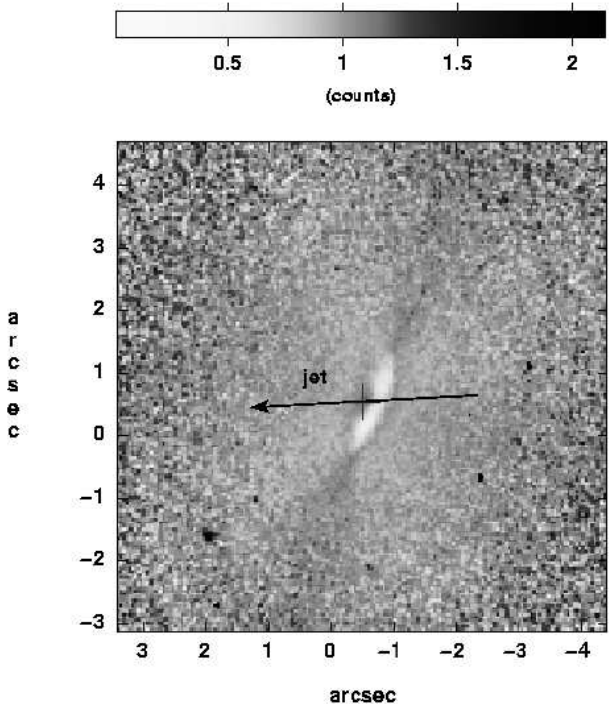,width=8cm,clip=}
\caption[]{Residual image of B2 0034+25; the arrow indicates the direction of the main jet}
\label{fig:b003425}
\end{figure}

\begin{figure}
%\picplace{6.0cm}
%\vspace{8cm}
\epsfig{file=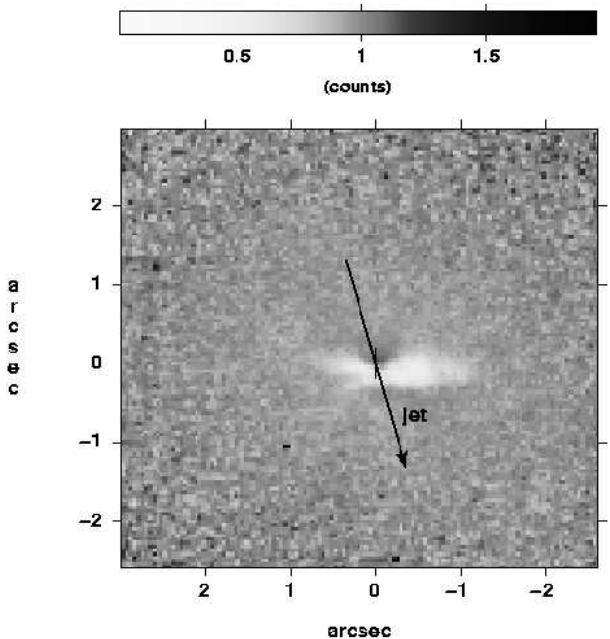,width=8cm,clip=}
\caption[]{Residual image of B2 1322+36; the arrow indicates the direction of the main jet}
\label{fig:b132236}
\end{figure}

The dust features were classified as disks, lanes (presumably edge-on disks), filaments (irregularly curved features, possibly
eccentric with respect to the galaxy nucleus), or complex (patchy structures). The classification of structures can be checked
with the help of the album of images given in Capetti et al. (\cite{capetti00}).
In Figs. \ref{fig:b003425} and \ref{fig:b132236} we give two of the best examples of disk-like structures found.
These residual images (or "ratio maps") were analyzed using the software package SYNAGE++ recently
developed by Murgia (\cite{murgia}). The arrows give the direction of the main jet. White colors
represent absorption and therefore dust features.

By no means all objects show absorption features: almost half (42 \%) of the observed galaxies have
a very smooth light distribution without any sign of circum-nuclear dust. When dust is seen
it is in most cases in the form of lanes or disks. Only in a minority of the
B2 galaxies do we see more complicated structures, like distorted filaments. A good example of
complex dust structures is B2 1350+31 (see Capetti et al. \cite{capetti00}).

\section{Discussion}
\label{sec:discussion}

\subsection{The detection of dust}
\label{subs:detection}

We tried to make quantitative estimates of the amount of dust present, by considering a positive
detection of dust only if in nine pixels that form a single area in the  
{\it ratio} maps the pixel
value was less than 0.85, (see also de Koff et al. \cite{dekoff00}).
We calculated the masses of the clumpy dust features in the usual way, following Sadler \& Gerhard
(\cite{sadler85}) and de Koff et al. (\cite{dekoff00}), by determining the covering factor of
the dust; this was done by summing all
pixels in the {\it ratio} maps which had a value $<$0.85, such that the dust mass is:
$M_{\rm dust} = \Sigma\times A_{\lambda}\times \Gamma^{-1}$, where $\Sigma$ is the area covered by
the dust, $A_{\lambda}$ the mean absorption of the area under consideration and $\Gamma$ the mass
absorption coefficient, taken as $6\times 10^{-6}$mag kpc$^2M_{\sun}^{-1}$ (see de Koff et al.
\cite{dekoff00}). The mass is based on the absorption in the $V$ images (see Sect. \ref{sec:presenceofdust}).
The mean visual absorption is in the range 0.2--0.5 mag for all objects.
The observed properties of the dust features (morphology, mean absorption and total area) 
are given in Table~\ref{tab:dustdata}. Since 3C sources already had been analyzed by
de Koff et al.~(\cite{dekoff00}), we did not repeat the calculation of $A_\lambda$, $\Sigma$ and
this is accounted for by an asterisk in the respective columns of Table~\ref{tab:dustdata}.
A non-detection of dust means that less than
nine pixels were found with value $<0.85$ in a single area. Corresponding upper limits are based on
this. In Table~\ref{tab:nodust} we list the upper limits to the dust masses
for those sources in which no dust was detected.

\begin{figure}
%\picplace{6.0cm}
%\vspace{8cm}
\epsfig{file=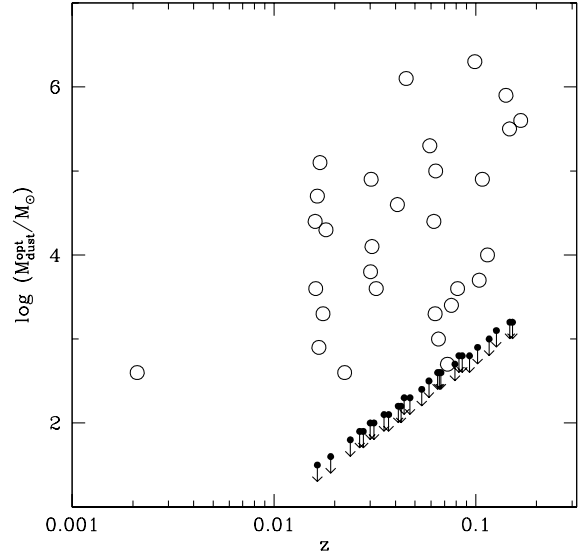,width=8cm,clip=}
\caption[]{Dust masses as a function of redshift; upper limits are based on the requirement that
at least 9 pixels in a contiguous area should have a value $<$0.85 in the residual image}
\label{fig:z_mo}
\end{figure}

We should remark that the detection of dust lanes or disks around galactic nuclei suffers from a
number of selection effects, which are not so easily quantifiable. First,  if the circum-nuclear
dust takes, as we believe, the form of disks, then its detection will clearly depend on orientation
of the disk with respect to the line of sight. Dust disks will be more easily seen if they are close
to edge-on (in which case they will be called "lanes"); up till now the only counter example in the
B2 sample is B2 0104+32 (3C 31), in which we are able to see fairly complete ellipses of dust around the nucleus.
Another obvious selection effect is due to redshift: since we require a minimum of nine pixels with
significant absorption in a contiguous area, the detectable dust mass depends on redshift (see
Sect. \ref{subs:dustmass}).  This effect is shown in Fig. \ref{fig:z_mo}, where we plot the dust
masses as
derived for B2 sources, including the upper limits, as a function of redshift. Is should also be
noted that the distribution of detected masses and upper limits appears to be discontinuous.

Using the criterion for detection we find that dust is frequently present:
30/57 (53 \%) of the galaxies show dust features, either in the form of bands or disk-like
structures, or more irregular patches. A summary of some general properties derived for these thirty sources
is given in Table \ref{tab:gendata}. Most columns of the table are obvious; it should only be
mentioned that we give two determinations of dust masses; the first, in Col. 6, is the mass
calculated from the absorption map as explained in this paper, while Col. 7 gives the dust mass
(or an upper limit) derived from the IRAS IR fluxes (if available).  A "J" in Col. 8 means that
the source has a radio jet. 
The percentage of dust detections in the B2 sample may be slightly higher than in the 3C sources,
which in 33 \% of the cases show dust (De Koff et al. \cite{dekoff00}). However, it does not appear that
there is any difference between FR I and FR II objects in the B2 sample: the FR I sources with dust
are 20/35 (57 \%), against 6/12 (50 \%) of the FR I-II and FR II sources (for the FR classification
see Capetti et al.~\cite{capetti00}).
Note that nine B2 sources are either compact nuclei or non-classifiable.

\begin{table*}
\caption[]{B2 radio galaxies: properties of dust}
\label{tab:dustdata}
\begin{flushleft}
\begin{tabular}{llrrrrr}
\hline\noalign{\smallskip}
Name & Dust & Major Axis & Minor Axis & PA (dust) & $A_{\lambda}$ & $\Sigma$ \\
     & type & kpc & kpc & deg. & mag & pixels \\
\noalign{\smallskip}
\hline\noalign{\smallskip}
0034+25 & disk     & 0.35 & 0.12 & 160 & 0.37 &   162 \\
0055+30 & disk     & 0.24 & 0.07 &  43 & 0.25 &   210 \\
0104+32 & disk     & 0.82 & 0.65 & 142 &  *   &    *  \\
0116+31 & filament & 5.02 & 0.08 &  20 & 0.42 &  2300 \\
0149+35 & disk     & 0.84 & 0.49 & 170 & 0.31 &  5087 \\
0648+27 & disk     & 0.28 & 0.16 &  45 & 0.40 &  1070 \\
0908+37 & disk     & 0.31 & 0.16 & 135 & 0.28 &    31 \\
0915+32 & disk     & 0.42 & 0.34 & 125 & 0.37 &   282 \\
1003+35 & disk     & 2.85 & 1.29 &  60 &  *   &    *  \\
1217+29 & complex  &    - &    - &   - & 0.42 &  3138 \\
1256+28 & disk     & 0.10 & 0.04 &   0 & 0.27 &    46 \\
1321+31 & disk     & 0.23 & 0.07 &  43 & 0.33 &   710 \\
1322+36 & disk     & 0.19 & 0.05 &  88 & 0.39 &   293 \\
1346+26 & filament & 4.02 &    - & 134 & 0.35 &  1307 \\
1339+26 & lane     & 0.19 & 0.05 & 175 & 0.28 &    29 \\
1347+28 & lane     & 0.36 &    - & 140 & 0.24 &     9 \\
1350+31 & complex  &    - &    - &   - &  *   &    *  \\
1357+28 & lane     & 0.33 & 0.08 &  77 & 0.23 &    35 \\
1430+25 & disk     & 0.21 & 0.09 &   0 & 0.28 &    41 \\
1447+27 & filament & 1.18 & 0.10 & 138 & 0.28 &   604 \\
1455+28 & filament & 4.59 & 1.68 & 158 & 0.48 &  1881 \\
1457+29 & lane     & 3.64 & 0.80 &  50 & 0.43 &   860 \\
1511+26 & lane     & 0.30 &    - & 120 &  *   &    *  \\
1525+29 & disk     & 0.43 &    - & 144 & 0.24 &    15 \\
1527+30 & disk     & 0.48 & 0.09 &   0 & 0.34 &    45 \\
1626+39 & filament &    - &    - &   - &  *   &    *  \\
1726+31 & lane     & 9.00 &    - & 127 &  *   &    *  \\
2116+26 & disk     & 0.44 & 0.12 &  65 & 0.37 & 10032 \\
2229+39 & disk     & 1.30 & 0.65 & 169 &  *   &    *  \\
2335+26 & disk     & 1.10 & 0.76 &   8 &  *   &    *  \\
\noalign{\smallskip}
\hline
\end{tabular}
\end{flushleft}
\end{table*}

\begin{table}
\caption[]{B2 radio galaxies: sources without dust}
\label{tab:nodust}
\begin{flushleft}
\begin{tabular}{llcr}
\hline\noalign{\smallskip}
Name & Redshift & $\log P/WHz^{-1}$ & $\log M/M_{\sun}$  \\
     &          & 1415 MHz          & upper limit        \\
\noalign{\smallskip}
\hline\noalign{\smallskip}
0055+26 & 0.0472 & 24.61 & 2.3 \\
0120+33 & 0.0164 & 22.30 & 1.5 \\
0708+32 & 0.0672 & 23.51 & 2.6 \\
0722+30 & 0.0191 & 22.77 & 1.6 \\
0755+37 & 0.0413 & 24.49 & 2.2 \\
0924+30 & 0.0266 & 23.52 & 1.9 \\
1003+26 & 0.1165 & 24.01 & 3.0 \\
1005+28 & 0.1476 & 24.25 & 3.2 \\
1101+38 & 0.0300 & 23.97 & 2.0 \\
1113+24 & 0.1021 & 23.65 & 2.9 \\
1204+34 & 0.0788 & 24.47 & 2.7 \\
1251+27 & 0.0857 & 25.37 & 2.8 \\
1257+28 & 0.0239 & 23.08 & 1.8 \\
1422+26 & 0.0370 & 24.00 & 2.1 \\
1450+28 & 0.1265 & 24.50 & 3.1 \\
1502+26 & 0.0540 & 25.36 & 2.4 \\
1512+30 & 0.0931 & 23.82 & 2.8 \\
1521+28 & 0.0825 & 24.58 & 2.8 \\
1553+24 & 0.0426 & 23.57 & 2.2 \\
1557+26 & 0.0442 & 22.81 & 2.3 \\
1610+29 & 0.0313 & 22.93 & 2.0 \\
1613+27 & 0.0647 & 24.03 & 2.6 \\
1615+32 & 0.1520 & 25.79 & 3.2 \\
1658+30 & 0.0351 & 23.88 & 2.1 \\
1827+32 & 0.0659 & 24.07 & 2.6 \\
1833+32 & 0.0586 & 25.07 & 2.5 \\
2236+35 & 0.0277 & 23.47 & 1.9 \\
\noalign{\smallskip}
\hline
\end{tabular}
\end{flushleft}
\end{table}

\begin{table*}
\caption[]{B2 radio galaxies: general properties}
\label{tab:gendata}
\begin{flushleft}
\begin{tabular}{lrcrrrrc}
\hline\noalign{\smallskip}
Name & $z$ & $\log P/WHz^{-1}$ & PA (radio) & $|\Delta PA|$ &
$\log M/M_{\sun}$ & $\log M/M_{\sun}$ & Jet \\
     &     & 1415 MHz          & deg        & deg            & 
dust        & IRAS \\
\noalign{\smallskip}
\hline\noalign{\smallskip}
0034+25 & 0.0321 & 23.20 &  93 & 67 & 3.6 & $<$6.3 & J \\
0055+30 & 0.0167 & 24.08 & 130 & 87 & 2.9 &    5.9 & J \\
0104+32 & 0.0169 & 24.21 & 160 & 18 & 5.1 &    6.4 & J \\
0116+31 & 0.0592 & 24.95 & 100 & 80 & 5.3 &    7.0 & - \\
0149+35 & 0.0160 & 22.33 &  65 & 75 & 4.4 &    6.2 & J \\
0648+27 & 0.0409 & 23.62 &   - &  - & 4.6 &    7.3 & - \\
0908+37 & 0.1040 & 24.84 &  15 & 60 & 3.7 & $<$7.2 & J \\
0915+32 & 0.0620 & 24.00 &  30 & 85 & 4.4 & $<$6.8 & J \\
1003+35 & 0.0989 & 25.78 & 120 & 60 & 6.3 & $<$6.6 & J \\
1217+29 & 0.0021 & 21.24 &   - &  - & 2.6 &    4.7 & J \\
1256+28 & 0.0224 & 23.05 &   - &  - & 2.6 & $<$5.4 & J \\
1321+31 & 0.0161 & 23.85 & 120 & 77 & 3.6 & $<$5.7 & J \\
1322+36 & 0.0175 & 24.55 &  15 & 73 & 3.3 & $<$5.8 & J \\
1346+26 & 0.0633 & 24.55 &  25 & 71 & 5.0 & $<$6.5 & - \\
1339+26 & 0.0757 & 24.30 &  30 & 35 & 3.4 & $<$7.0 & - \\
1347+28 & 0.0724 & 24.05 &  50 & 90 & 2.7 & $<$7.0 & J \\
1350+31 & 0.0452 & 25.03 &  90 &  - & 6.1 &    6.8 & J \\
1357+28 & 0.0629 & 24.03 &   0 & 77 & 4.4 & $<$6.8 & J \\
1430+25 & 0.0813 & 24.20 &  20 & 20 & 3.6 &      - & - \\
1447+27 & 0.0306 & 22.78 &  27 & 69 & 4.1 & $<$6.1 & - \\
1455+28 & 0.1411 & 25.22 &  38 & 60 & 5.9 &      - & - \\
1457+29 & 0.1470 & 24.89 & 155 & 75 & 5.5 &      - & - \\
1511+26 & 0.1078 & 25.34 &  20 & 80 & 4.9 &      - & - \\
1525+29 & 0.0653 & 23.98 &  30 & 66 & 3.0 & $<$6.8 & J \\
1527+30 & 0.1143 & 24.05 &  45 & 45 & 4.0 & $<$7.3 & - \\
1626+39 & 0.0303 & 24.49 &  90 &  - & 4.9 & $<$5.7 & J \\
1726+31 & 0.1670 & 25.89 & 110 & 17 & 5.6 & $<$7.6 & - \\
2116+26 & 0.0164 & 22.79 & 157 & 88 & 4.7 &    6.3 & J \\
2229+39 & 0.0181 & 24.03 &  10 & 21 & 4.3 &    6.3 & J \\
2335+26 & 0.0301 & 24.88 & 125 & 63 & 3.8 & $<$6.0 & J \\
\noalign{\smallskip}
\hline
\end{tabular}
\end{flushleft}
\end{table*}

Our data strengthen the findings by  Martel et al. (\cite{martel99}) who noted that disk-like
structures are preferentially associated to FR I type radio sources. Since FR I sources often have
detected radio jets, it is expected that the presence of disklike structures is also related to the
presence of jets. This is indeed the case, but we can even make a more detailed statement: if a
source has complex dust features, i.e. no disk or lane, then most of the sources have no radio jet
(7/9), while a majority of objects with a disk-like feature also have a radio jet (16/21). Another
way of saying this is that the large majority (16/18) of the B2 galaxies with a radio jet also have
a disk or lane (see Table \ref{tab:dustjet}). Fisher's exact probability test (see e.g. Siegel
\cite{siegel56}) gives a 
probability of 0.9~\% that the distribution in  Table \ref{tab:dustjet} is due to chance.

It should also be mentioned that if we limit ourselves to
sources with a jet and with redshift $z<0.03$ (like the objects discussed by Verdoes Kleijn et al.
\cite{verdoes}), then dust is detected even in 80 \% of the cases, against the 53 \% overall.
This increase in the detection of dust features in more nearby galaxies regards only sources with a
radio jet: if no jet is present the fraction of sources with dust is close to 50 \%, both for nearby
sources and for the sources with $z > 0.03$.

We find no connection between dust content and radio size: the sources with dust
have a median linear size of $94\pm 35$ kpc, against $76\pm 30$ for the sources  without dust.

\begin{table*}
\caption[]{Detection of dust and radio jets}
\label{tab:dustjet}
\begin{flushleft}
\begin{tabular}{lrrr}
\hline\noalign{\smallskip}
                                & Radio sources & Radio sources & Total \\
                                                                & with jet      & without jet \\
                                \noalign{\smallskip}
\hline\noalign{\smallskip}
Galaxies with dust-disk or lane & 16 &  5 & 21 \\
Galaxies with complex dust      &  2 &  7 &  9 \\
Total                           & 18 & 12 & 30 \\
\noalign{\smallskip}
\hline
\end{tabular}
\end{flushleft}
\end{table*}

\begin{table*}
\caption[]{Detection of dust and optical core}
\label{tab:dustcore}
\begin{flushleft}
\begin{tabular}{lrrr}
\hline\noalign{\smallskip}
                      & Galaxies with & Galaxies without & Total \\
                                          & optical core  & optical core \\
\noalign{\smallskip}
\hline\noalign{\smallskip}
Galaxies with dust    &  9 & 21 & 30 \\
Galaxies without dust &  9 & 18 & 27 \\
Total                 & 18 & 39 & 57 \\
\noalign{\smallskip}
\hline
\end{tabular}
\end{flushleft}
\end{table*}

Table \ref{tab:dustcore} gives the relation between the detection of dust and the presence
of an optical core.
The data on the optical cores were taken from Capetti et al. (\cite{capetti02}). It is
quite obvious that there is no
significant preference for sources with dust to have no detected optical core, or, reading the table
differently, for galaxies without optical core to be more dusty. Although one would expect that the
possibility to detect an optical core is negatively influenced by the presence of dust, in line with
the usual picture of the optical core being surrounded by a dusty disk or torus which obscures the
core when the disk is seen close to edge-on, the absence of any significant effect may be influenced
by the strong selection effects discussed earlier: the detection of dust and the optical core in the
standard model both depend on the orientation with respect to the observer. Especially the detection
of dust may be sensitive, in a complicated way, to the redshift of the galaxy as well.
However, perhaps most important of all, Chiaberge, Capetti \& Celotti (\cite{chiaberge99}) concluded
that the optical nucleus is very often not obscured even if dust is present, so that the dust must be
optically thin. This provides a natural explanation for the absence of a correlation between dust
and optical nucleus. In fact, all objects present a mean absorption consistent
with an optical depth $\tau < 1$.

\subsection{Orientation of dust features and the direction of the radio axis}
\label{subs:orientation}
\begin{figure}
%\picplace{6.0cm}
%\vspace{8cm}
\epsfig{file=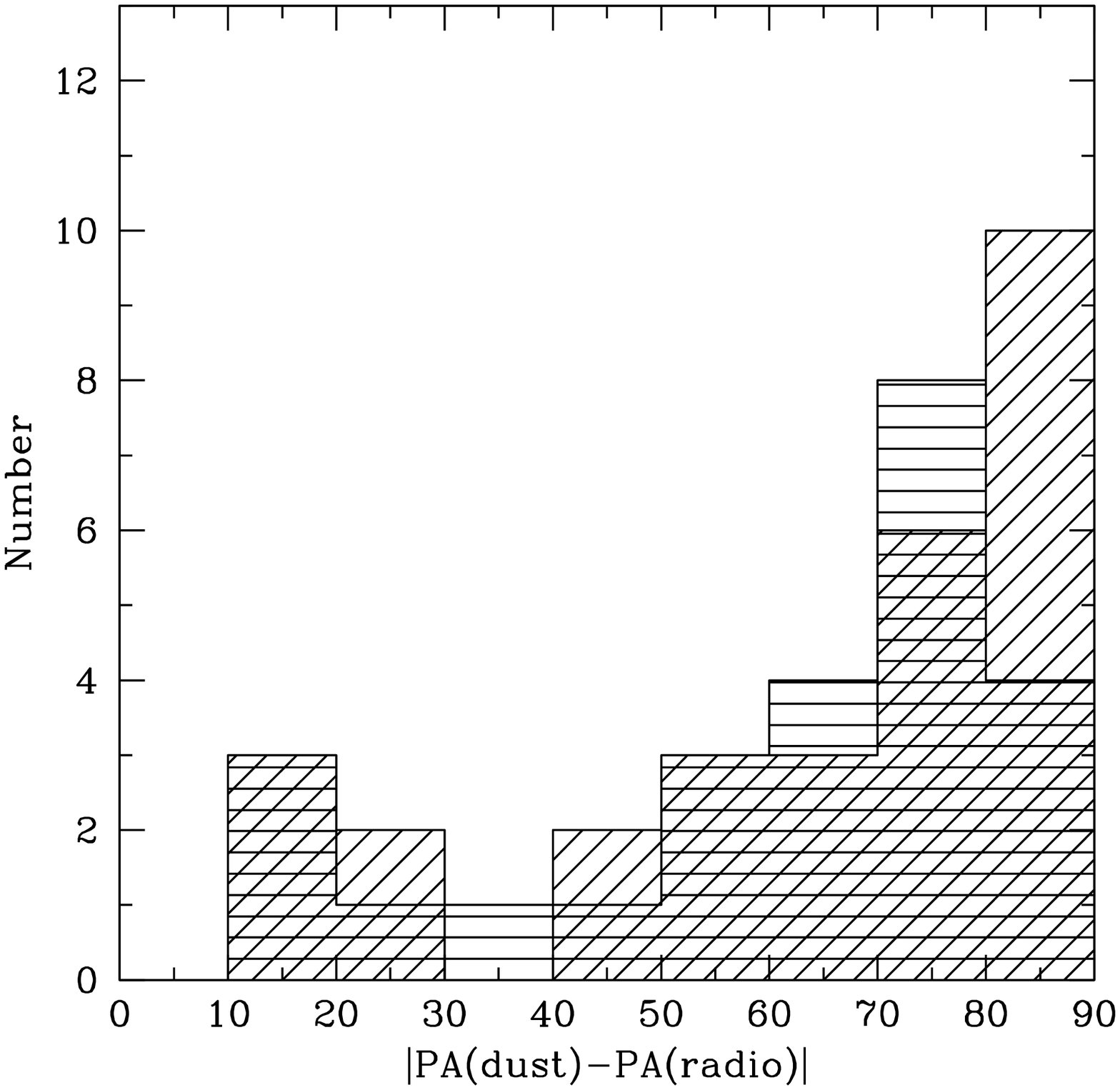,width=8cm,clip=}
\caption[]{Radio-dust alignments in B2 sources (horizontally shaded) and 3C sources that are
not part of the B2 sample (diagonally shaded)}
\label{fig:dhist}
\end{figure}

\begin{figure}
%\picplace{6.0cm}
%\vspace{8cm}
\epsfig{file=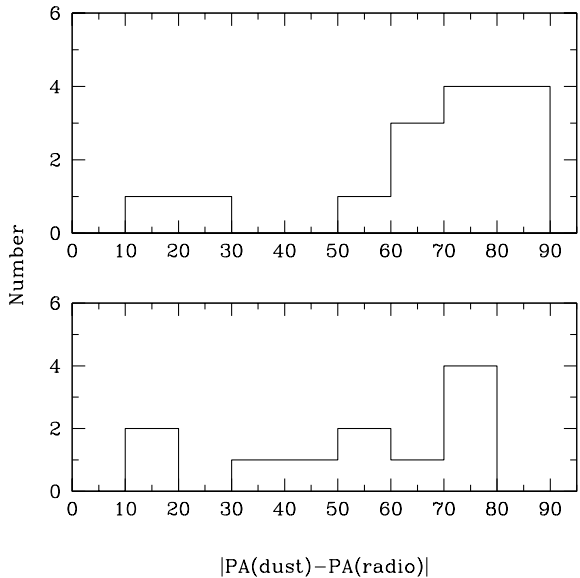,width=8cm,clip=}
\caption[]{Radio-dust alignments according to the presence (upper panel) or absence (lower
panel) of a radio jet}
\label{fig:dpa_jet}
\end{figure}

Alignment effects involving the radio axis and some optical axis (for example the major axis of the
galaxy) have been searched for in the past, but the results were mostly inconclusive. However in
nearby galaxies harbouring a (weak) radio source an alignment between the central dust feature and the
radio axis was found (Kotanyi \& Ekers \cite{kotanyi}).
Naively one would expect that the directions of the radio axis and the disk or lane are strongly
correlated. Indeed, there is a general tendency for the radio axis to be perpendicular to the disk or
lane, as can be seen in Fig. \ref{fig:dhist}.  There may also be a tendency for a more
prominent alignment if a radio jet is detected (see Fig. \ref{fig:dpa_jet}), however, the
effect, if it exists at all, is weak.

\begin{figure}
%\picplace{6.0cm}
%\vspace{8cm}
\epsfig{file=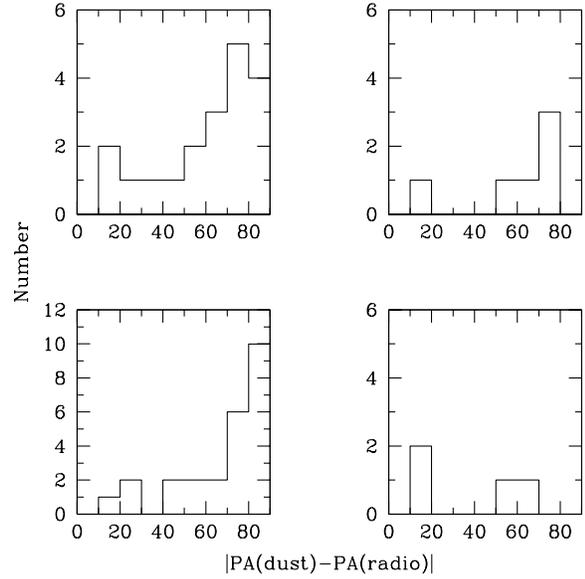,width=8cm,clip=}
\caption[]{Alignments according to dust structure. For definition of simple and complex dust
structure see text. Upper left: simple B2 sources; upper right:
complex B2; lower left: simple 3C; lower right: complex 3C}
\label{fig:dcom_sim}
\end{figure}
\begin{figure}
%\picplace{6.0cm}
%\vspace{8cm}
\epsfig{file=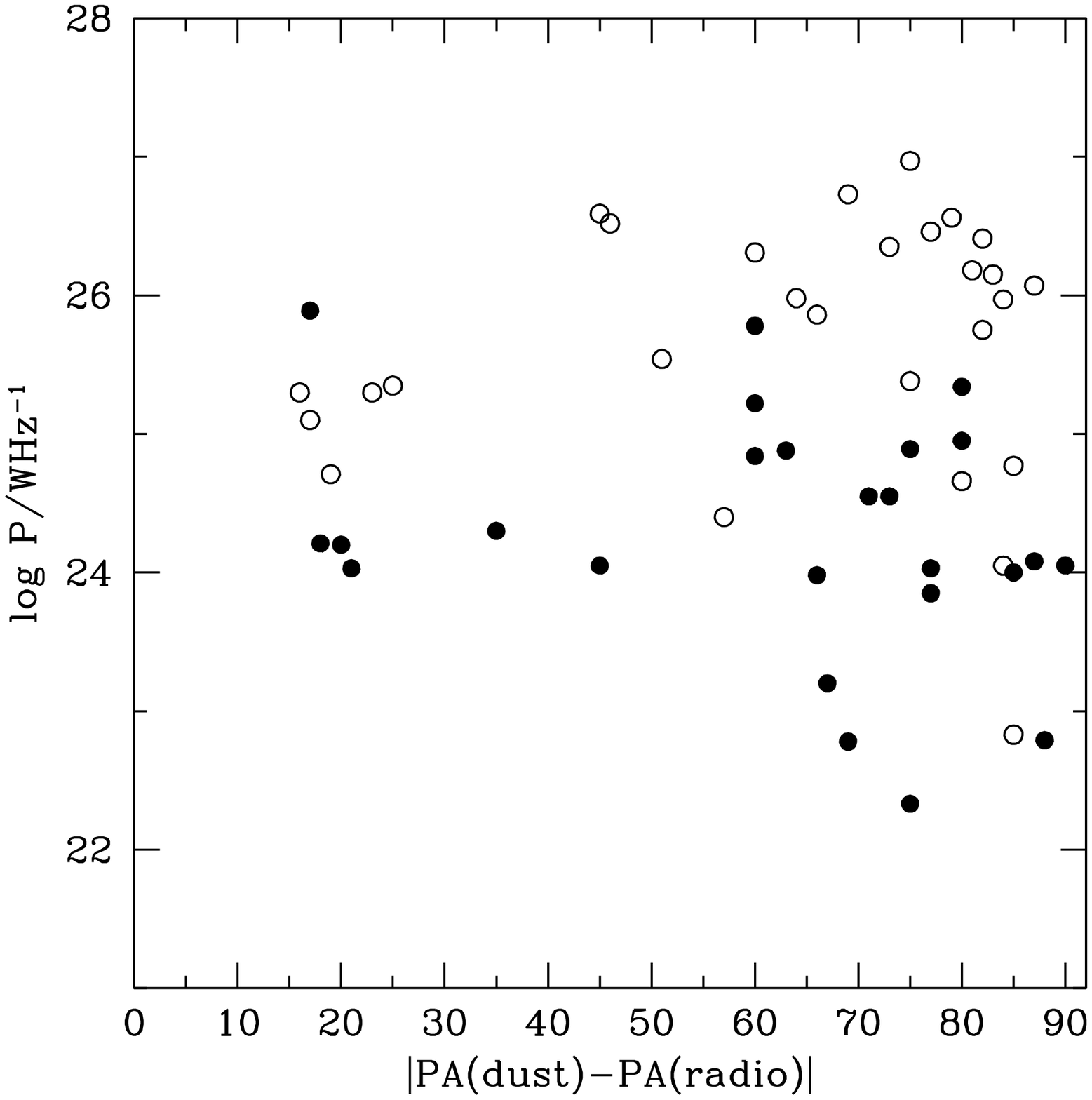,width=8cm,clip=}
\caption[]{Alignments according to radio power. Filled symbol: B2; open symbol: 3C sources}
\label{fig:dpa_p}
\end{figure}

According to de Koff et al. (\cite{dekoff96}), who used recent HST images of 3C sources, the
alignment is most pronounced in sources with (small) disk structures.
Figure~\ref{fig:dcom_sim} partly  confirms this statement. Indeed the difference between simple (i.e.
disk-like) and more complex structures in non-B2 3C galaxies (lower panel) is significant at a confidence level
of 98.5~\%. However in our data on the B2 galaxies (upper panel) no significant difference is seen (confidence level $<90~\%$). 
The mean alignment angles are $62\degr $ for both simple and complex dust features in B2 sources, and $68\degr $ and
$38\degr $ for simple and complex 3C sources respectively.
In this and other figures we have
defined disklike structures (including straight lanes) as simple, and all other dust features
(including distorted lanes) as complex. 

As we remarked in the previous Section, the simple structures tend to be found in sources with radio
jets. It is well known that  jets are more easily detected in weaker radio sources of type FR I and it is therefore to
be expected that dusty disks and lanes are more frequently present in the weaker radio sources with
jets; any alignment between radio axis and dust-disks should be sought there.
This is
illustrated in Fig. \ref{fig:dpa_p}, where we plot the alignment angle against radio power.
Clearly many sources, and actually all of the weak ones with  $P\leq 10^{24}$WHz$^{-1}$, have an alignment angle 
$\geq 60\degr $. For the FR II sources (coinciding mostly with the non-B2 3C
sources in Fig. \ref{fig:dpa_p}) there are quite a few sources with small angles ($<60\degr $).
The mean alignment angle is $75.3\pm 8.7 \degr $ for the B2 sources with $P\leq 10^{24}$WHz$^{-1}$, 
against $57.3\pm 25\degr $ for the stronger B2 sources. The difference between the average angles
is not significant, but the variance of the distributions is different (confidence level $>99$~\%
from Fisher's F-distribution). This of course reflects the fact that all the weak sources are 
clustered towards high alignment angles.

\subsection{Dust masses}
\label{subs:dustmass}
\begin{figure}
%\picplace{6.0cm}
%\vspace{8cm}
\epsfig{file=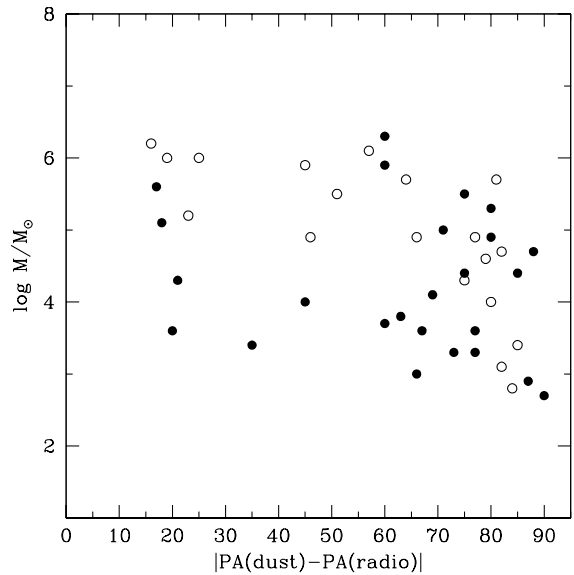,width=8cm,clip=}
\caption[]{Alignments according to dustmass. Filled symbol: B2; open symbol: 3C sources}
\label{fig:dpa_m}
\end{figure}

As explained in Sect. 4.1, we calculated the masses of the dust features in the usual way,
following Sadler \& Gerhard (\cite{sadler85}). In Fig. \ref{fig:dpa_m} we show the derived
dust masses against the alignment angle. No correlation is seen.

\begin{figure}
%\picplace{6.0cm}
%\vspace{8cm}
\epsfig{file=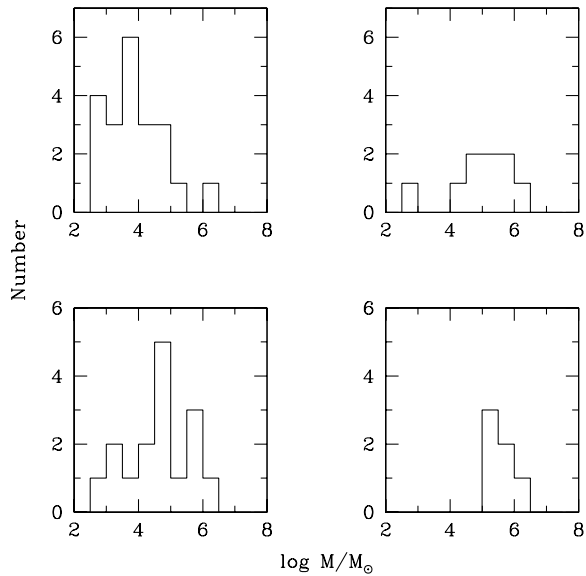,width=8cm,clip=}
\caption[]{Masses according to dust structure (see text).
Upper panels: B2, lower panels: 3C sources. Left:
simple structures; right: complex structures}
\label{fig:mcom_sim}
\end{figure}

A confirmation that the simple structures have on average lower masses is shown in Fig. \ref{fig:mcom_sim}: the
lower masses correspond to the disk and lane structures. 
For the B2 galaxies with simple dust structures we find an average $<\log (M/M{\sun}>=3.9$, with a standard 
deviation of the mean of 0.20, while complex structures have an average of 5.0 with standard deviation 0.36. According to
Student's t-test this difference is significant at the 99.5~\% level. A similar difference is obtained for the 3C galaxies:
$<\log (M/M{\sun}>=4.68\pm 0.25$  for simple dust and $5.2\pm 0.13$ for complex dust. This difference
is significant at a confidence level of 98~\%.
The higher mass values
of the 3C galaxies reflect the fact that they are generally stronger radio sources.
Anyway, it is possible that the objects
classified as containing complex dust structures also have small disks, hidden in the complexity of the
dust features.

\begin{figure}
%\picplace{6.0cm}
%\vspace{8cm}
\epsfig{file=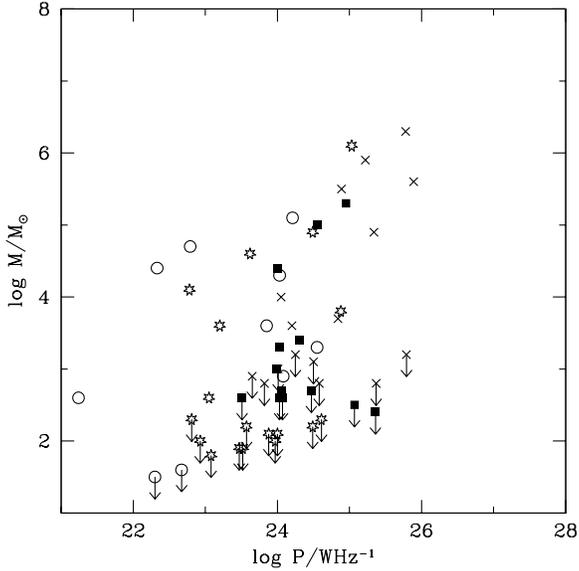,width=8cm,clip=}
\caption[]{Dust mass as a function of total radio power (B2 sources). Different symbols
indicate different redshift intervals. Open circles: $z<0.02$; open stars: $0.02< z < 0.05$; 
filled squares: $0.05 < z < 0.08$; crosses: $z > 0.08$}
\label{fig:pmz}
\end{figure}

Most striking, however, is the correlation shown in Fig. \ref{fig:pmz}, between the total radio
power and the {\it detected} dust masses.  A linear regression between $\log P$ and 
$\log M/M_{\sun}$ gives a slope of $0.61\pm 0.16$.
We can reject (using the recipe given in e.g. Haber \& Runyon \cite{haber73}) the hypothesis that
radio power and dust mass are uncorrelated at the 99.9 \% level. 
However, it is known that radio
power and redshift are correlated in flux limited samples; using the test discussed by Macklin
(\cite{macklin82}), we can reject the hypothesis that the apparent correlation 
between power and dust mass is due exclusively to a correlation between P and z and M and z, at the
level of $4.3 \sigma$. We may illustrate this strong correlation (independent of redshift), 
by dividing our
sample in four redshift intervals ($z<0.02$, $0.02<z<0.05$, $0.05<z<0.08$ and $z>0.08$). 
The correlation remains in three of the four redshift interval  (see Fig. \ref{fig:pmz}); 
although it is formally not significant in the lowest redshift bin ($0.45 \sigma$) the correlation
is very strong indeed in the other redshift bins, even if the number of objects involved is
only of the order of 7 or 8 per bin: the significance levels are 3.1, 4.3 and 4.9 $\sigma$ in order of increasing
redshift. The location of the upper limits in Fig. \ref{fig:pmz} strongly suggests that the
distribution in the $\log P - \log M$ plane is bimodal, that is, if there is dust detected in the form of
disks or lanes, then the amount of dust depends on radio power (or vice-versa), but in addition to this
there are radio sources in which no dust is detected, practically independently of the radio power
(the slow rising of the upper limits in Fig. \ref{fig:pmz} with radio power is due to increasing
redshift). 

No correlation is found between detected dust masses and the absolute magnitude of the
galaxy and this is in line with the result of de Koff et al. (\cite{dekoff00}); 
therefore the relation dust mass against power stands on its own, and does not appear
to be induced by other possible correlations between galaxy properties.

\begin{figure}
%\picplace{6.0cm}
%\vspace{8cm}
\epsfig{file=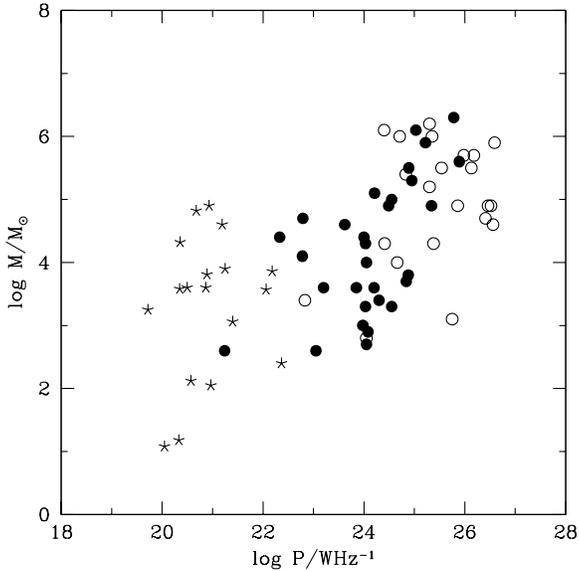,width=8cm,clip=}
\caption[]{Dust masses as a function of total radio power, for B2 (filled dots), 3C (open
circles) and sources from Tran et al. (stars)}
\label{fig:pm_new}
\end{figure}

In order to compare the correlation derived from the B2 sample with other data we show in Fig. \ref{fig:pm_new}
the same plot as in Fig.~\ref{fig:pmz}, but now with non-B2 3C sources added and also an optically selected sample of nearby 
($z < 0.011$) E and S0 galaxies described in Tran et al. (\cite{tran01}). For clarity no upper limits were plotted.
The galaxies in which both radio emission (based on the NVSS) and dust were detected are 18 of the
107 galaxies discussed by Tran et al. (\cite{tran01}). Since the selection criteria of 
the Tran et al. (\cite{tran01}) are rather complicated no statistical conclusions should be drawn
from the location of these galaxies in Fig. \ref{fig:pm_new}; we just remark that they roughly
fall in the region were we would expect them to be according to the correlation based on the B2
sample.

\begin{figure}
%\picplace{6.0cm}
%\vspace{8cm}
\epsfig{file=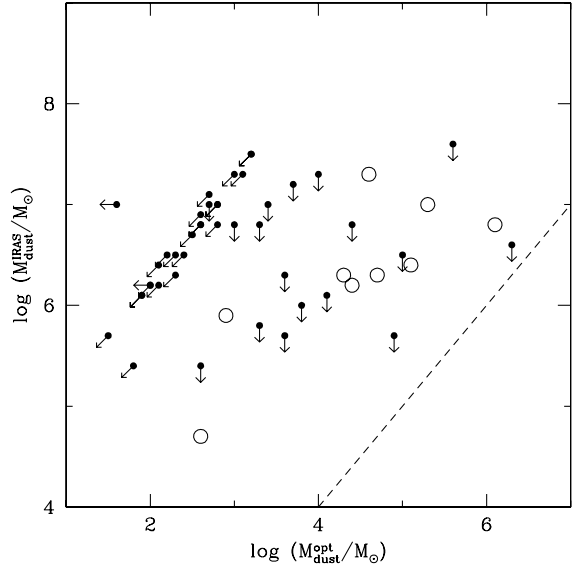,width=8cm,clip=}
\caption[]{The relation between dust masses derived from absorption, as calculated here and from
IRAS observations }
\label{fig:mo_mir}
\end{figure}

In a number of cases we can also derive dust masses from IRAS data, although most B2 sources have
only upper limits. We use the 100 $\mu$m data given by Impey \& Gregorini (\cite{impey}),
if available. The dust masses were calculated using a dust temperature $T_{\rm d} = 30$\ K (see de Koff
et al. \cite{dekoff00});
the results are shown in Fig. \ref{fig:mo_mir}. The derived IRAS masses are typically a
factor $\sim$100 higher than the masses derived from our absorption analysis. To bring the masses
down to the level of the masses determined here from the absorption analysis, we would need to
increase the dust temperature in most cases to unrealisticly high values ($>>100$\ K). Only for
B2 0104+32 (3C 31), B2 1350+31 and B2 2116+26 can we bring the dust masses in agreement with $T_{\rm d} < 100$\ K. We
therefore conclude that the {\it detected} IRAS dust masses are in general one or two orders of magnitude higher
than the masses derived from the absorption analysis. The upper limits in Fig.~\ref{fig:mo_mir} are consistent with this.
It is of course not surprising that the IRAS masses are larger, as the IRAS fluxes measure the overall presence of dust (including diffuse components),
while the absorption analysis is only sensitive to sharp and well defined features, which in many
cases means only the dusty disk close to the nucleus. Apparently the circumnuclear features
measured by us represent not much more than about between 1 and 10 \% of the overall dust.

\section{Summary}
\label{sec:summary}
\begin{enumerate}
\item Dust features are detected in slightly over 50 \% (30/57) of the B2 sources, which is a
bit higher than was found for 3C sources by de Koff et al. (\cite{dekoff00}).
Although this would suggest a difference between FRI (majority of B2) and FRII (majority in 3C)
sources no significant effect is found.
\item If a B2 radio source has jets, then in 16/18 sources the dust around the galactic nucleus comes in the
form of disks or lanes, i.e. of simples structures; this correlation is very strong: the probability that
the distribution is only due to chance is 0.9~\% . Conversely, in seven of the nine sources with complex dust structures no radio jet
is seen.
\item The radio jets tend to be perpendicular to the dust lanes or disks, in particular
the weaker FR I radio sources, with $P<10^{24}$WHz$^{-1}$. Moreover small dust masses
(with $\log M/M_{\sun} < 3$) are found mostly in these weaker sources. In part this may be a selection
effect, because the weakest sources are also the most nearby, in which small dust masses are more
easily detected.
\item We have shown that if circum-nuclear dust is present in low luminosity radio sources, then its
amount and its morphology are linked to the properties of the radio sources. In other words the radio
source "knows" about the presence of dust. We base this conclusion on the fact that there is a well
established correlation between the dust mass and the total (or core) radio power, as described in
the previous Section. Since the radio and optical properties are (in observational sense) only
indirectly dependent via the distance, the only way to explain this correlation away is through a
selection effect that works both on the derived dust masses and on the radio power. The fact that the
correlation between radio power and dust mass remains valid also within different redshift intervals
strongly suggests that the correlation is intrinsic. However this correlation holds only if a
significant amount of dust is present, and there may be a discontinuity (or bimodality in the mass
distribution), as is also clear from Fig. \ref{fig:pmz}.
\end{enumerate}

\begin{acknowledgements}
This research has made use of the NASA/IPAC Extragalactic Database (NED) which is operated by the
Jet Propulsion Laboratory, California Institute of Technology, under contract with the National
Aeronautics and Space Administration.
\end{acknowledgements}

\end{document}